\documentclass{article}
\usepackage{spconf,amsmath,graphicx}

\usepackage{multirow}
\usepackage{graphicx}
\usepackage{amssymb, amsmath, bm, mathtools,array}
\usepackage{textcomp}
\usepackage{hyperref}
\usepackage{verbatim,lipsum}

\RequirePackage[normalem]{ulem} %DIF PREAMBLE
\RequirePackage{color}\definecolor{BLUE}{rgb}{0,0.20,0.75} %DIF PREAMBLE
\RequirePackage{color}\definecolor{BROWN}{RGB}{60,128,49} %DIF PREAMBLE
\newcommand{\bs}[1]{\boldsymbol{#1}}
\title{Neural source-filter-based waveform model for statistical parametric speech synthesis}
\name{Xin Wang$^1$, Shinji Takaki$^1$, Junichi Yamagishi$^{1}$\sthanks{This work was partially supported by JST CREST Grant Number JPMJCR18A6, Japan and by MEXT KAKENHI Grant Numbers (16H06302, 16K16096, 17H04687, 18H04120, 18H04112, 18KT0051), Japan.}}
\address{$^1$National Institute of Informatics, Japan \\
  {\small \tt wangxin@nii.ac.jp, takaki@nii.ac.jp, jyamagis@nii.ac.jp}}

\begin{document}
\ninept
\maketitle
%%%%%%%%%%%%%%
\begin{abstract}
Neural waveform models such as the WaveNet are used in many recent text-to-speech systems, but the original WaveNet is quite slow in waveform generation because of its autoregressive (AR) structure. Although faster non-AR models were recently reported, they may be prohibitively complicated due to the use of a distilling training method and the blend of other disparate training criteria. This study proposes a non-AR neural source-filter waveform model that can be directly trained using spectrum-based training criteria and the stochastic gradient descent method. Given the input acoustic features, the proposed model first uses a source module to generate a sine-based excitation signal and then uses a filter module to transform the excitation signal into the output speech waveform. Our experiments demonstrated that the proposed model generated waveforms at least 100 times faster than the AR WaveNet and the quality of its synthetic speech is close to that of speech generated by the AR WaveNet. Ablation test results showed that both the sine-wave excitation signal and the spectrum-based training criteria were essential to the performance of the proposed model. 

\end{abstract}

%%%%%%%%%%%%%%
\section{Introduction}
\label{sec:intro}
Text-to-speech (TTS) synthesis, a technology that converts texts into speech waveforms, 
has been advanced by using end-to-end architectures \cite{shen2018natural} and 
neural-network-based waveform models \cite{oord2016wavenet,pmlr-v80-oord18a, ping2018clarinet}. 
Among those waveform models, the WaveNet \cite{oord2016wavenet} directly models the 
distributions of waveform sampling points and has demonstrated outstanding performance.
The vocoder version of WaveNet \cite{Tamamori2017}, which converts the acoustic features into the waveform, 
also outperformed other vocoders for the pipeline TTS systems \cite{wangICASSP2018}. 

As an autoregressive (AR) model, the WaveNet is quite slow in waveform generation because it 
has to generate the waveform sampling points one by one. 
%Furthermore, its complicated structure is difficult to interpret.
To improve the generation speed, the Parallel WaveNet \cite{pmlr-v80-oord18a} and the ClariNet \cite{ping2018clarinet}
introduce a distilling method to transfer `knowledge' from a teacher AR WaveNet to a student non-AR model that
simultaneously generates all the waveform sampling points. However, the concatenation of two large models and 
the mix of distilling and other training criteria reduce the model interpretability and raise 
the implementation cost. 

In this paper, we propose a neural source-filter waveform model that converts acoustic features into speech waveforms.
Inspired by classical speech modeling methods \cite{hedelin1981tone, mcaulay1986speech}, 
we used a source module to generate a sine-based excitation signal with a specified fundamental frequency (F0).
We then used a dilated-convolution-based filter module to transform the sine-based excitation into the speech waveform. 
The proposed model was trained by minimizing spectral amplitude and phase distances, which can be efficiently implemented using discrete Fourier transforms (DFTs). 
Because the proposed model is a non-AR model, it generates waveforms much faster than the AR WaveNet. 
A large-scale listening test showed that the proposed model was close to the AR WaveNet in terms of the Mean opinion score (MOS) on the quality of synthetic speech. 
An ablation test showed that both the sine-wave excitation and the spectral amplitude distance were crucial  to the proposed model.

The model structure and training criteria are explained in Section~\ref{sec:model_method}, after which
the experiments are described in Section~\ref{sec:exp}. Finally, this paper is summarized and concluded in Section~\ref{sec:con}.

\begin{figure*}[!t]
\includegraphics[width=\textwidth]{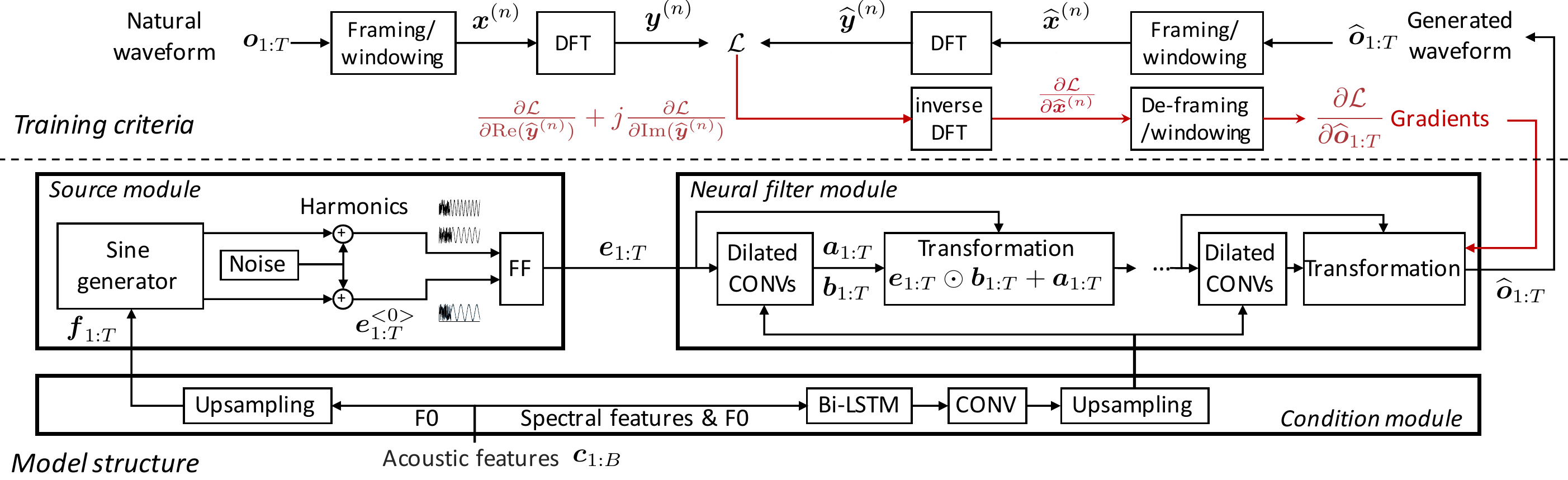}
\vspace{-6mm}
\caption{Structure of proposed model. $B$ and $T$ denote lengths of input feature sequence and output waveform, respectively. {FF}, {CONV}, and {Bi-LSTM} denote feedforward, convolutional, and bi-directional recurrent layers, respectively. DFT denotes discrete Fourier transform.}
\label{fig_model}
\end{figure*}

\section{Proposed model and training criteria}
\label{sec:model_method}
\subsection{Model structure}
\label{sec:model}
The proposed model (shown in Figure~\ref{fig_model}) converts an input acoustic feature sequence $\bs{c}_{1:B}$ of length $B$ into a speech waveform $\widehat{\bs{o}}_{1:T}$ of length $T$.
It includes a source module that generates an excitation signal $\bs{e}_{1:T}$, a filter module that transforms $\bs{e}_{1:T}$ into the speech waveform, and a condition module that processes the acoustic features for the source and filter modules. 
None of the modules takes the previously generated waveform sample as the input. 
The waveform is assumed to be real-valued, i.e., $\widehat{o}_t\in\mathbb{R}, 0<{t}\leq{T}$.

\subsubsection{Condition module}
\label{sec:model_cond}
The condition module takes as input the acoustic feature sequence $\bs{c}_{1:B}=\{\bs{c}_1, \cdots, \bs{c}_B\}$, where each $\bs{c}_{b}=[f_b, \bs{s}_b^{\top}]^{\top}$ contains the F0 $f_b$ and the spectral features $\bs{s}_b$ of the $b$-th speech frame. 
The condition module upsamples the F0 by duplicating $f_b$ to every time step within the $b$-th frame and 
feeds the upsampled F0 sequence $\bs{f}_{1:T}$ to the source module. 
Meanwhile, it processes $\bs{c}_{1:B}$ using a bi-directional recurrent layer with long-short-term 
memory (LSTM) units \cite{Graves2008} and a convolutional (CONV) layer, after which 
the processed features are upsampled and sent to the filter module. 
The LSTM and CONV were used so that the condition module was similar to that of the WaveNet-vocoder \cite{weko_185771_1} in the experiment.
They can be replaced with a feedforward layer in practice.

\subsubsection{Source module}
\label{sec:model_source}
Given the input F0 sequence $\bs{f}_{1:T}$, the source module generates a sine-based excitation signal 
${\bs{e}_{1:T}}=\{e_1,\cdots, e_T\}$, where ${e}_t\in\mathbb{R}, \forall{t}\in\{1,\cdots, T\}$.
Suppose the F0 value of the $t$-th time step is $f_t\in\mathbb{R}_{\geq{0}}$, and $f_t=0$ denotes being unvoiced. 
By treating $f_t$ as the instantaneous frequency \cite{carson1937variable}, a signal $\bs{e}_{1:T}^{<0>}$ can be generated as
\begin{align}
{e}_t^{<0>} = \begin{dcases}
\alpha\sin(\sum_{k=1}^{t}2\pi\frac{{f}_k}{N_s} + {\phi}) + {n}_t, &\text{if }{f_t}>0 \\ 
\frac{1}{3\sigma} {n}_t, & \text{if } f_t = 0\\ 
\end{dcases},
\label{eq:sine}
\end{align}
where ${n}_t \sim \mathcal{N}(0, \sigma^2)$ is a Gaussian noise, $\phi\in[-\pi, \pi]$ is a random initial phase,
and $N_s$ is equal to the waveform sampling rate. 

%For unvoiced steps, ${e}_t^{(0)}$ is simply set as ${e}_t^{(0)} = \frac{1}{3\sigma} {n}_t$. 
%where $\frac{1}{3\sigma}$ ensures that dynamic range of the noise is roughly equal to that of the sine signal. 
%\begin{equation}
%{e}_t^{(0)} = \text{sin}(\sum_{k=1}^{t}2\pi{f}_k\frac{k}{N_s} + {\phi}) + {n}_t,
%\label{eq:sine}
%\end{equation}

Although we can directly set $\bs{e}_{1:T}=\bs{e}_{1:T}^{<0>}$, 
we tried two additional tricks.
First, a `best' phase $\phi^{*}$ for $\bs{e}_{1:T}^{<0>}$ can be determined in the training stage 
by maximizing the correlation between $\bs{e}_{1:T}^{<0>}$ and the natural waveform $\bs{o}_{1:T}$. 
During generation, $\phi$ is randomly generated. 
The second method is to generate harmonics by increasing $f_k$ in Equation~($\ref{eq:sine}$) and use a feedforward (FF) layer to 
merge the harmonics and $\bs{e}_{1:T}^{<0>}$ into $\bs{e}_{1:T}$. 
In this paper we use 7 harmonics and set $\sigma=0.003$ and $\alpha=0.1$.

\subsubsection{Neural filter module}
\label{sec:model_filter}
Given the excitation signal ${\bs{e}_{1:T}}$ from the source module and the processed acoustic features from the condition module, 
the filter module modulates ${\bs{e}_{1:T}}$ using multiple stages of dilated convolution and affine transformations similar to those in ClariNet \cite{ping2018clarinet}. 
For example, the first stage takes ${\bs{e}_{1:T}}$ and the processed acoustic features as input 
and produces two signals $\bs{a}_{1:T}$ and $\bs{b}_{1:T}$ using dilated convolution. 
The ${\bs{e}_{1:T}}$ is then transformed using ${\bs{e}_{1:T}}\odot{\bs{b}_{1:T}} + \bs{a}_{1:T}$, 
where $\odot$ denotes element-wise multiplication.
The transformed signal is further processed in the following stages, and the output of the final stage
is used as generated waveform $\widehat{\bs{o}}_{1:T}$.

The dilated convolution blocks are similar to those in Parallel WaveNet \cite{pmlr-v80-oord18a}. 
Specifically, each block contains multiple dilated convolution layers with a filter size of 3. 
The outputs of the convolution layers are merged with the features from the condition module
through gated activation functions \cite{pmlr-v80-oord18a}. After that, the merged features are transformed into 
$\bs{a}_{1:T}$ and $\tilde{\bs{b}}_{1:T}$.  To make sure that $\bs{b}_{1:T}$ is positive, 
$\bs{b}_{1:T}$ is parameterized as $\bs{b}_{1:T}=\exp(\tilde{\bs{b}}_{1:T})$.

Unlike ClariNet or Parallel WaveNet, the proposed model does not use the distilling method. It is unnecessary to compute 
the mean and standard deviation of the transformed signal. 
Neither is it necessary to form the convolution and transformation blocks as an inverse autoregressive flow \cite{NIPS2016_6581}. 

\subsection{Training criteria in frequency domain}
Because speech perception heavily relies on acoustic cues in the frequency domain, 
we define training criteria that minimize the spectral amplitude and phase distances, which can be implemented using DFTs.
Given these criteria, the proposed model is trained using the stochastic gradient descent (SGD) method. 
%The matrix form of the criteria and gradients can be found in another work \cite{dnn_tsnet}.

\subsubsection{Spectral amplitude distance}
Following the convention of short-time Fourier analysis, 
we conduct waveform framing and windowing before producing the spectrum of each frame. 
For the generated waveform $\widehat{\bs{o}}_{1:T}$, we use $\widehat{\bs{x}}^{(n)}=[\widehat{x}^{(n)}_1, \cdots, \widehat{x}^{(n)}_M]^{\top}\in\mathbb{R}^{M}$ to denote the $n$-th waveform frame of length $M$. 
We then use $\widehat{\bs{y}}^{(n)}=[\widehat{y}^{(n)}_1, \cdots, \widehat{y}^{(n)}_K]^{\top}\in\mathbb{C}^{K}$ to
denote the spectrum of $\widehat{\bs{x}}^{(n)}$ calculated using $K$-point DFT.
We similarly define $\bs{x}^{(n)}$ and $\bs{y}^{(n)}$ for the natural waveform $\bs{o}_{1:T}$.

Suppose the waveform is sliced into $N$ frames. Then the log spectral amplitude distance is defined as follows:
\begin{equation}
\mathcal{L}_s = \frac{1}{2}\sum_{n=1}^{N}\sum_{k=1}^{K}\Big[\log\frac{\texttt{Re}(y_k^{(n)})^2+\texttt{Im}(y_k^{(n)})^2}{\texttt{Re}(\widehat{y}_k^{(n)})^2+\texttt{Im}(\widehat{y}_k^{(n)})^2}\Big]^2,
\label{eq:dft_spectral}
\end{equation}
where $\texttt{Re}(\cdot)$ and $\texttt{Im}(\cdot)$ denote the real and imaginary parts of a complex number, respectively. 

Although $\mathcal{L}_s$ is defined on complex-valued spectra, the gradient 
$\frac{\partial{\mathcal{L}_s}}{\partial{\widehat{\bs{o}}_{1:T}}}\in\mathbb{R}^{T}$ for SGD training can be efficiently calculated. 
Let us consider the $n$-th frame and compose a complex-valued vector $\bs{g}^{(n)} = \frac{\partial{\mathcal{L}_s}}{\partial{\texttt{Re}(\widehat{\bs{y}}^{(n)})}} + j \frac{\partial{\mathcal{L}_s}}{\partial{\texttt{Im}(\widehat{\bs{y}}^{(n)})}}\in\mathbb{C}^{K}$, where the $k$-th element is ${g}^{(n)}_k = \frac{\partial{\mathcal{L}_s}}{\partial{\texttt{Re}(\widehat{{y}}_k^{(n)})}} + j \frac{\partial{\mathcal{L}_s}}{\partial{\texttt{Im}(\widehat{{y}}_k^{(n)})}}\in\mathbb{C}$.
It can be shown that, as long as $\bs{g}^{(n)}$ is Hermitian symmetric, the inverse DFT of $\bs{g}^{(n)}$ is equal to $\frac{\partial\mathcal{L}_s}{\partial\widehat{\bs{x}}^{(n)}}=[
\frac{\partial\mathcal{L}_s}{\partial\widehat{{x}}_1^{(n)}},\cdots,\frac{\partial\mathcal{L}_s}{\partial\widehat{{x}}_m^{(n)}},\frac{\partial\mathcal{L}_s}{\partial\widehat{{x}}_M^{(n)}}]\in\mathbb{R}^{M}$ \footnote{In the implementation using fast Fourier transform, $\widehat{\bs{x}}^{(n)}$ of length $M$ is zero-padded to length $K$ before DFT. Accordingly, the inverse DFT of $\bs{g}^{(n)}$ also gives the gradients w.r.t. the zero-padded part, which should be discarded (see \url{https://arxiv.org/abs/1810.11946}).}.
Using the same method, $\frac{\partial\mathcal{L}_s}{\partial\widehat{\bs{x}}^{(n)}}$ for ${n}\in\{1,\cdots,N\}$ can be computed in parallel. 
Given $\{\frac{\partial\mathcal{L}_s}{\partial\widehat{\bs{x}}^{(1)}}, \cdots, \frac{\partial\mathcal{L}_s}{\partial\widehat{\bs{x}}^{(N)}}\}$, 
the value of each $\frac{\partial{\mathcal{L}_s}}{\partial{\widehat{{o}}_{t}}}$ in $\frac{\partial{\mathcal{L}_s}}{\partial{\widehat{\bs{o}}_{1:T}}}$ can be easily accumulated since the relationship between $\widehat{{o}}_t$ and each $\widehat{x}^{(n)}_{m}$ has been determined by the framing and windowing operations. 

In fact, $\frac{\partial{\mathcal{L}_s}}{\partial{\widehat{\bs{o}}_{1:T}}}\in\mathbb{R}^{T}$
can be calculated in the same manner no matter how we set the framing and DFT configuration, i.e., 
the values of $N$, $M$, and $K$.
Furthermore, multiple $\mathcal{L}_s$s with different configurations can be computed, 
and the gradients $\frac{\partial{\mathcal{L}_s}}{\partial{\widehat{\bs{o}}_{1:T}}}$ can be simply summed up. 
For example, using the three $\mathcal{L}_{s}$s in Table~\ref{tab:dft_config} was found to be essential to the proposed model (see Section~\ref{sec:ex_ablation}).

The Hermitian symmetry of $\bs{g}^{(n)}$ is satisfied if $\mathcal{L}_s$ is carefully defined. For example, $\mathcal{L}_s$ can be the square error or Kullback-Leibler divergence (KLD) of the spectral amplitudes \cite{lee2001algorithms,Takaki2017}. The phase distance defined below also satisfies the requirement.

\begin{table}[!t]
\caption{Three framing and DFT configurations for $\mathcal{L}_{s}$ and $\mathcal{L}_{p}$}
\vspace{-3mm}
\begin{center}
\begin{tabular}{rccc}
\hline\hline
 & $\mathcal{L}_{s1}$\&$\mathcal{L}_{p1}$ & $\mathcal{L}_{s2}$\&$\mathcal{L}_{p2}$ & $\mathcal{L}_{s3}$\&$\mathcal{L}_{p3}$ \\
 \hline
DFT bins $K$         &  512 & 128 & 2048 \\
Frame length $M$  & 320 (20 ms) & 80 (5 ms)  & 1920 (120 ms) \\
Frame shift         & 80 (5 ms)   & 40 (2.5 ms) & 640 (40 ms) \\
\hline\hline
\multicolumn{4}{c}{Note: all configurations use Hann window. } \\
\end{tabular}
\end{center}
\label{tab:dft_config}
\vspace{-7mm}
\end{table}%

\subsubsection{Phase distance}
Given the spectra, a phase distance \cite{dnn_tsnet} is computed as
\begin{equation}
\begin{split}
&\mathcal{L}_p= \frac{1}{2}\sum_{n=1}^{N}\sum_{k=1}^{K}\Big|1-\exp(j(\widehat\theta^{(n)}_k - {\theta}^{(n)}_k))\Big|^2\\
&=\sum_{n=1}^{N}\sum_{k=1}^{K}\Big[1-\frac{\texttt{Re}(\widehat{y}_k^{(n)})\texttt{Re}({y}_k^{(n)})+\texttt{Im}(\widehat{y}_k^{(n)})\texttt{Im}({y}_k^{(n)})}{|\widehat{y}_k^{(n)}||{y}_k^{(n)}|}\Big]
\label{eq:dft_phase}
\end{split},
\end{equation}
where $\widehat\theta^{(n)}_k$ and $\theta^{(n)}_k$ are the phases of $\widehat{y}^{(n)}_k$ and ${y}^{(n)}_k$, respectively. 
The gradient $\frac{\partial{\mathcal{L}_p}}{\partial{\widehat{\bs{o}}_{1:T}}}$ can be computed by the same procedure as 
$\frac{\partial{\mathcal{L}_s}}{\partial{\widehat{\bs{o}}_{1:T}}}$. 
Multiple $\mathcal{L}_p$s and $\mathcal{L}_s$s with different framing and DFT configurations can be 
added up as the ultimate training criterion $\mathcal{L}$.
For different $\mathcal{L}_{*}$s, additional DFT/iDFT and framing/windowing blocks should be added to 
the model in Figure~\ref{fig_model}.

\section{Experiments}
\label{sec:exp}
\subsection{Corpus and features}
This study used the same Japanese speech corpus and data division recipe as our previous study \cite{luong2018investigating}. 
This corpus \cite{kawai2004ximera} contains neutral reading speech uttered by a female speaker. 
Both validation and test sets contain 480 randomly selected utterances.
Among the 48-hour training data,  9,000 randomly selected utterances (15 hours) were used 
as the training set in this study. For the ablation test in Section~\ref{sec:ex_ablation}, the training set
was further reduced to 3,000 utterances (5 hours).
Acoustic features, including 60 dimensions of Mel-generalized cepstral coefficients (MGCs) \cite{to1994mel} and 1 
dimension of F0, were extracted from the 48 kHz waveforms at a frame shift of 5 ms using WORLD \cite{morise2016world}. 
The natural waveforms were then downsampled to 16 kHz for model training and the listening test.

\subsection{Comparison of proposed model, WaveNet, and WORLD}
\label{sec:ex_compare}

The first experiment compared the four models listed in Table~\ref{tab:models}\footnote{The models were implemented using a modified CURRENNT toolkit \cite{weninger2015introducing} on a single P100 Nvidia GPU card. Codes, recipes, and generated speech can be found on \url{https://nii-yamagishilab.github.io}.}.
\begin{comment}
\begin{itemize}
\item \texttt{NSF}: the proposed neural source-filter waveform model;
\item \texttt{WAD}: a WaveNet-vocoder using a categorical distribution on 10-bits quantized $\mu$-law compressed waveform;
\item \texttt{WAC}: a WaveNet-vocoder using a Gaussian distribution on float-valued waveform;
\item \texttt{WOR}: the WORLD vocoder;
\end{itemize}
\end{comment}
The \texttt{WAD} model, which was trained in our previous study \cite{wangICASSP2018}, contained a condition module, a post-processing module, and 40 dilated CONV blocks, where the $k$-th CONV block had a dilation size of $2^{\text{modulo}(k,10)}$. 
\texttt{WAC} was similar to \texttt{WAD} but used a Gaussian distribution to model the raw waveform at the output layer \cite{ping2018clarinet}. 

The proposed \texttt{NSF} contained 5 stages of dilated CONV and transformation, each stage including 10 convolutional layers 
with a dilation size of $2^{\text{modulo}(k,10)}$ and a filter size of 3. 
Its condition module was the same as that of \texttt{WAD} and \texttt{WAC}. 
\texttt{NSF} was trained using $\mathcal{L} = \mathcal{L}_{s1}+\mathcal{L}_{s2}+\mathcal{L}_{s3}$,  and the configuration of each $\mathcal{L}_{s*}$ is listed in Table~\ref{tab:dft_config}. 
The phase distance $\mathcal{L}_{p*}$ was not used in this test. 

\begin{table}[!t]
\caption{Models for comparison test in Section~\ref{sec:ex_compare}}
\vspace{-3mm}
\begin{center}
\begin{tabular}{cl}
\hline\hline
\texttt{WOR} & WORLD vocoder \\
\texttt{WAD} & WaveNet-vocoder for 10-bit discrete $\mu$-law waveform \\
\texttt{WAC} & WaveNet-vocoder using Gaussian dist. for raw waveform  \\
\texttt{NSF} &  Proposed model  for raw waveform \\
\hline\hline
\end{tabular}
\end{center}
\label{tab:models}
\vspace{-6mm}
\end{table}

\begin{figure}[!t]
\includegraphics[width=\columnwidth]{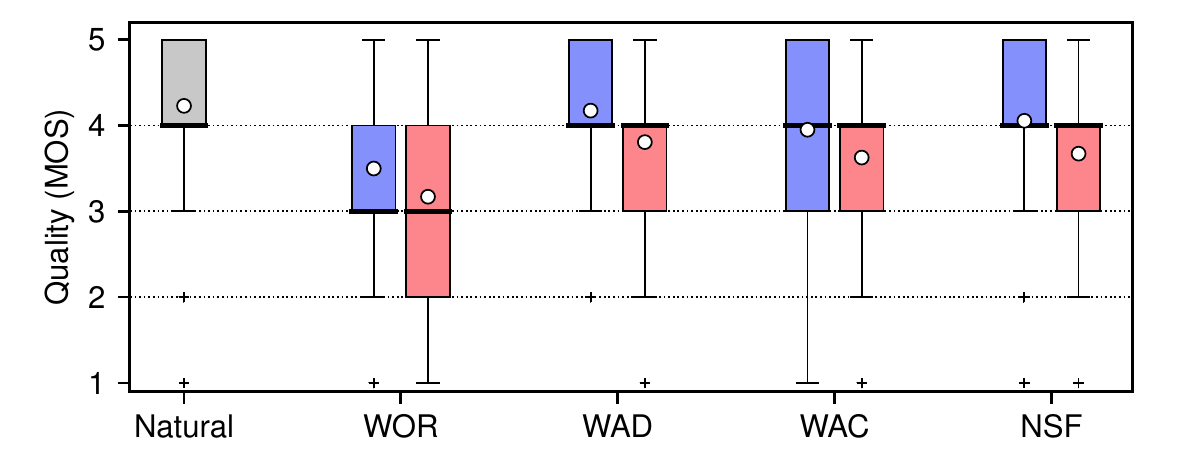}
\vspace{-8mm}
\caption{MOS scores of natural speech,  synthetic speech given natural acoustic features (\textcolor{blue}{blue}), and {synthetic speech given acoustic features generated from acoustic models (\textcolor{red}{red})}. White dots are mean values.}
\label{fig:mos_1}
\vspace{-2mm}
\end{figure}

Each model generated waveforms using natural and generated acoustic features, where
the generated acoustic features were produced by the acoustic models in our previous study \cite{wangICASSP2018}. 
The generated and natural waveforms were then evaluated by paid native Japanese speakers.
In each evaluation round the evaluator listened to one speech waveform in each screen and rated the speech quality on a 1-to-5 MOS scale. 
The evaluator can take at most 10 evaluation rounds and can replay the sample during evaluation.
The waveforms in an evaluation round were for the same text and were played in a random order.
Note that the waveforms generated from \texttt{NSF} and \texttt{WAC} were converted to 16-bit PCM format before evaluation.

\begin{table}[t!]
\vspace{-2mm}
\caption{Average number of waveform points generated in 1 s}
\begin{center}
\begin{tabular}{ccc}
\hline\hline
  \texttt{WAD} & \texttt{NSF} (memory-save mode) & \texttt{NSF} (normal mode)\\
  \hline
0.19k & 20k & 227k  \\
\hline\hline
\end{tabular}
\vspace{-7mm}
\end{center}
\label{tab:speed}
\end{table}

A total of 245 evaluators conducted 1444 valid evaluation rounds in all, and the results are plotted in Figure~\ref{fig:mos_1}.
Two-sided Mann-Whitney tests showed that the difference between any pair of models is statistically significant ($p<0.01$)
except \texttt{NSF} and \texttt{WAC} when the two models used generated acoustic features. 
In general, \texttt{NSF} outperformed \texttt{WOR} and \texttt{WAC} but performed slightly worse than  \texttt{WAD}.
The gap of the mean MOS scores between \texttt{NSF} and \texttt{WAD} was about 0.12, given either natural or generated acoustic features. 
A possible reason for this result may be the difference between the non-AR and AR model structures,
which is similar to the difference between the finite and infinite impulse response filters. 
\texttt{WAC} performed worse than \texttt{WAD} because some syllables were perceived to be trembling in pitch,
which may be caused by the random sampling generation method. 
\texttt{WAD} alleviated this artifact by using a one-best generation method in voiced regions \cite{wangICASSP2018}.

\begin{figure*}[!t]
\includegraphics[width=\textwidth]{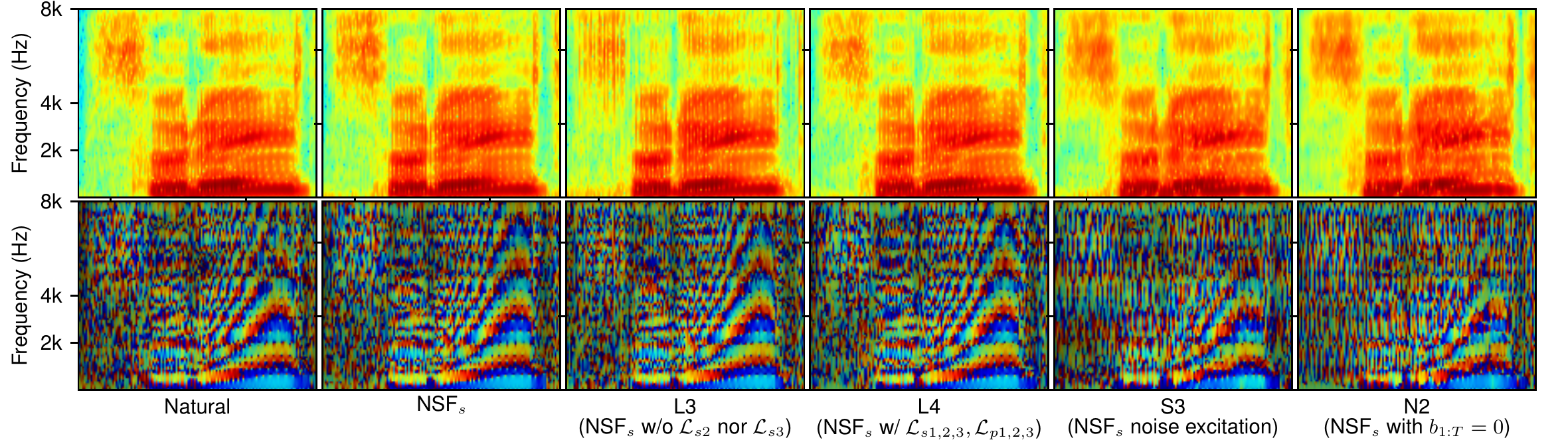}
\vspace{-9mm}
\caption{Spectrogram (top) and instantaneous frequency (bottom) of natural waveform and  waveforms generated from models in Table~\ref{tab:models_2} given natural acoustic features in test set (utterance AOZORAR\_03372\_T01). Figures are plotted using 5 ms frame length and 2.5 ms frame shift.}
\vspace{-5mm}
\label{fig:spec}
\end{figure*}

After the MOS test, we compared the waveform generation speed of \texttt{NSF} and \texttt{WAD}.
The implementation of \texttt{NSF} has a normal generation mode and a memory-save one. 
The normal mode allocates all the required GPU memory once but cannot generate waveforms 
longer than 6 seconds because of the insufficient memory space in a single GPU card.
The memory-save mode can generate long waveforms because it releases and allocates the memory 
layer by layer, but the repeated memory operations are time consuming.

We evaluated \texttt{NSF} using both modes on a smaller test set,
in which each of the 80 generated test utterances was around 5 seconds long. 
As the results in Table~\ref{tab:speed} show, \texttt{NSF} is much faster than \texttt{WAD}. 
Note that \texttt{WAD} allocates and re-uses a small size of GPU memory, which needs no
repeated memory operation. \texttt{WAD} is slow mainly because of the AR generation process.
Of course, both \texttt{WAD} and \texttt{NSF} can be improved if our toolkit is further optimized.
Particularly, if the memory operation can be sped up, the memory-save mode of 
\texttt{NSF} will be much faster.

\subsection{Ablation test on proposed model}
\label{sec:ex_ablation}
This experiment was an ablation test on \texttt{NSF}. 
Specifically, the 11 variants of \texttt{NSF} listed in Table~\ref{tab:models_2} were trained using the 5-hour training set. 
For a fair comparison, \texttt{NSF} was re-trained using the 5-hour data, and this variant is referred to as \texttt{NSF}$_{\text{s}}$. 
The speech waveforms were generated given the natural acoustic features and rated in 1444 evaluation rounds 
by the same group of evaluators in Section~\ref{sec:ex_compare}. This test excluded natural waveform for evaluation.

\begin{comment}
\begin{itemize}
\item \raggedright{
\textcolor{black}{On $\mathcal{L}$:} \texttt{L1} didn't use $\mathcal{L}_{s2}$; \newline
\textcolor{white}{On $\mathcal{L}$:} \texttt{L2} didn't use $\mathcal{L}_{s3}$; \newline 
\textcolor{white}{On $\mathcal{L}$:} \texttt{L3} didn't use $\mathcal{L}_{s2}$ nor $\mathcal{L}_{s3}$; \newline 
\textcolor{white}{On $\mathcal{L}$:} \texttt{L4} used all three $\mathcal{L}_{s}$ and $\mathcal{L}_{p}$; \newline 
\textcolor{white}{On $\mathcal{L}$:} \texttt{L5} used all three $\mathcal{L}_{s}$ of spectral amplitude KLD; 
}
\item \raggedright{
\textcolor{black}{On source:} \texttt{S1} didn't use harmonics;  \newline 
\textcolor{white}{On source:} \texttt{S2} didn't use harmonics nor phase match;  \newline 
\textcolor{white}{On source:} \texttt{S3} only used noise as excitation
}
\item \raggedright{
\textcolor{black}{On filter:} \texttt{N1} set ${\bs{b}_{1:T}}=1$ in each transformation layer; \newline
\textcolor{white}{On filter:} \texttt{N2} set ${\bs{b}_{1:T}}=0$ in each transformation layer;
}
\item \raggedright{
\textcolor{black}{On distilling:} \texttt{D1} was trained using three $\mathcal{L}_{s}$ and distilling.  
Student was initialized by \texttt{S3}, and teacher was fixed to \texttt{WAC}.}
\end{itemize}
\end{comment}

\begin{table}[!t]
\vspace{-2mm}
\caption{Models for ablation test (Section~\ref{sec:ex_ablation})}
\vspace{-3mm}
\begin{center}
\begin{tabular}{cl}
\hline\hline
\texttt{NSF}$_{s}$ & \texttt{NSF} trained on 5-hour data \\
\hline
\texttt{L1} & \texttt{NSF}$_{s}$ without using $\mathcal{L}_{s3}${\textcolor{white}{ssssssssss}} (i.e., $\mathcal{L} = \mathcal{L}_{s1}+\mathcal{L}_{s2}$) \\
\texttt{L2} & \texttt{NSF}$_{s}$ without using $\mathcal{L}_{s2}${\textcolor{white}{ssssssssss}} (i.e., $\mathcal{L} = \mathcal{L}_{s1}+\mathcal{L}_{s3}$)\\
\texttt{L3} & \texttt{NSF}$_{s}$ without using $\mathcal{L}_{s2}$ nor  $\mathcal{L}_{s3}${\textcolor{white}{sssssssss}} (i.e., $\mathcal{L} = \mathcal{L}_{s1}$) \\
\texttt{L4} & \texttt{NSF}$_{s}$ using $\mathcal{L} = \mathcal{L}_{s1}+\mathcal{L}_{s2} + \mathcal{L}_{s3} + \mathcal{L}_{p1} + \mathcal{L}_{p2} +\mathcal{L}_{p3}$ \\
\texttt{L5} & \texttt{NSF}$_{s}$ using KLD of spectral amplitudes\\
\hline
\texttt{S1} & \texttt{NSF}$_{s}$ without harmonics \\
\texttt{S2} & \texttt{NSF}$_{s}$ without harmonics or `best' phase $\phi^{*}$ \\
\texttt{S3} & \texttt{NSF}$_{s}$ only using noise as excitation \\
\hline
\texttt{N1} & \texttt{NSF}$_{s}$ with ${\bs{b}_{1:T}}=1$ in filter's transformation layers \\
\texttt{N2} & \texttt{NSF}$_{s}$ with ${\bs{b}_{1:T}}=0$ in filter's transformation layers \\
%\hline
%\texttt{D1} & \texttt{S3} with distilling given teacher \texttt{WAC} \\
\hline\hline
\end{tabular}
\vspace{-7mm}
\end{center}
\label{tab:models_2}
\end{table}

\begin{figure}[!t]
\includegraphics[width=\columnwidth]{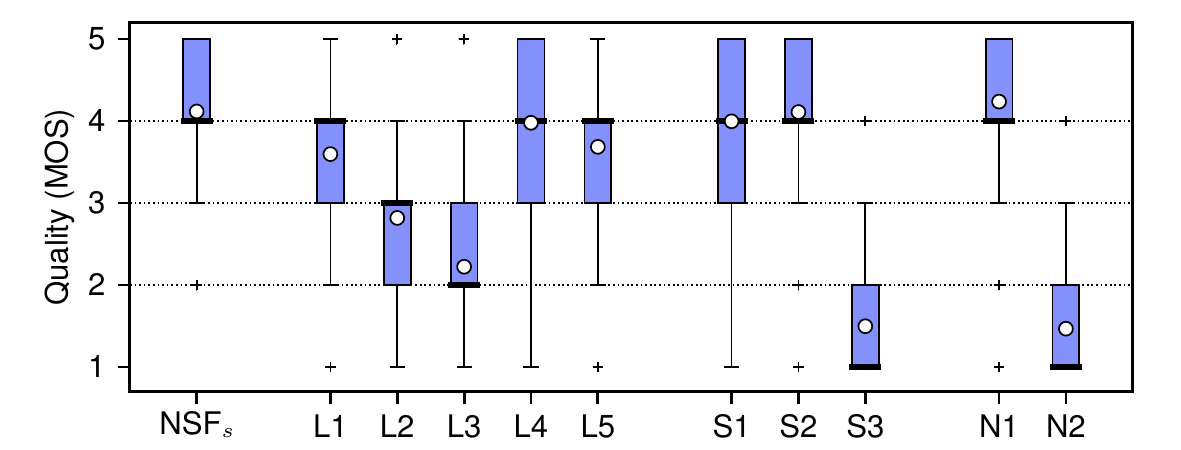}
\vspace{-8mm}
\caption{MOS scores of synthetic samples from \texttt{NSF}$_s$ and its variants given natural acoustic features. White dots are mean MOS scores.}
\label{fig:mos_2}
\vspace{-4mm}
\end{figure}

The results are plotted in Figure~\ref{fig:mos_2}. 
The difference between \texttt{NST}$_s$ and any other model  except \texttt{S2} was statistically significant ($p<0.01$).
Comparison among \texttt{NST}$_s$, \texttt{L1}, \texttt{L2}, and \texttt{L3} shows that using multiple 
$\mathcal{L}_{s}$s listed in Table~\ref{tab:dft_config} is beneficial. For \texttt{L3} that used only $\mathcal{L}_{s1}$, 
the generated waveform points clustered around one peak in each frame, and the waveform suffered from a pulse-train noise. 
This can be observed from \texttt{L3} of Figure~\ref{fig:spec}, whose spectrogram in the high frequency band shows more clearly vertical strips than other models. 
Accordingly, this artifact can be alleviated by adding $\mathcal{L}_{s2}$ with a frame length of 5 ms for model training, which explained the improvement in \texttt{L1}. Using phase distance (\texttt{L4}) didn't improve the speech quality even though the value of the phase distance was consistently decreased on both training and validation data.

The good result of \texttt{S2} indicates that a single sine-wave excitation with a random initial phase also works. Without the sine-wave excitation, \texttt{S3} generated waveforms that were intelligible but lacked stable harmonic structure. \texttt{N1} slightly outperformed \texttt{NSF}$_s$ while \texttt{N2}  produced unstable harmonic structures. Because the transformation in \texttt{N1} is equivalent to skip-connection \cite{he2016deep}, the result indicates that the skip-connection may help the model training.

\section{Conclusion}
\label{sec:con}
In this paper, we proposed a neural waveform model with separated source and filter modules.
The source module produces a sine-wave excitation signal with a specified F0, and the filter module uses dilated convolution to transform the excitation into a waveform. 
Our experiment demonstrated that the sine-wave excitation was essential for generating waveforms with harmonic structures. We also found that multiple spectral-based training criteria and the transformation in the filter module contributed to the performance of the proposed model.
Compared with the AR WaveNet, the proposed model generated speech with a similar quality at a much faster speed.

The proposed model can be improved in many aspects. 
For example, it is possible to simplify the dilated convolution blocks. 
It is also possible to try classical speech modeling methods, including glottal waveform excitations \cite{fant1985four, juvela2016high},
two-bands or multi-bands approaches \cite{makhoul1978mixed,Griffin-multiband} on waveforms. 
When applying the model to convert linguistic features into the waveform, we observed the over-smoothing affect in the high-frequency band
and will investigate the issue in the future work.

%%%%%%%%%%%%%%
\vfill\pagebreak

% References should be produced using the bibtex program from suitable
% BiBTeX files (here: strings, refs, manuals). The IEEEbib.bst bibliography
% style file from IEEE produces unsorted bibliography list.
% -------------------------------------------------------------------------
\bibliographystyle{IEEEbib}
\bibliography{BIB}

\appendix
\onecolumn
\section{Forward computation}
\begin{figure*}[h]
\centering
\fbox{\includegraphics[width=0.95\textwidth]{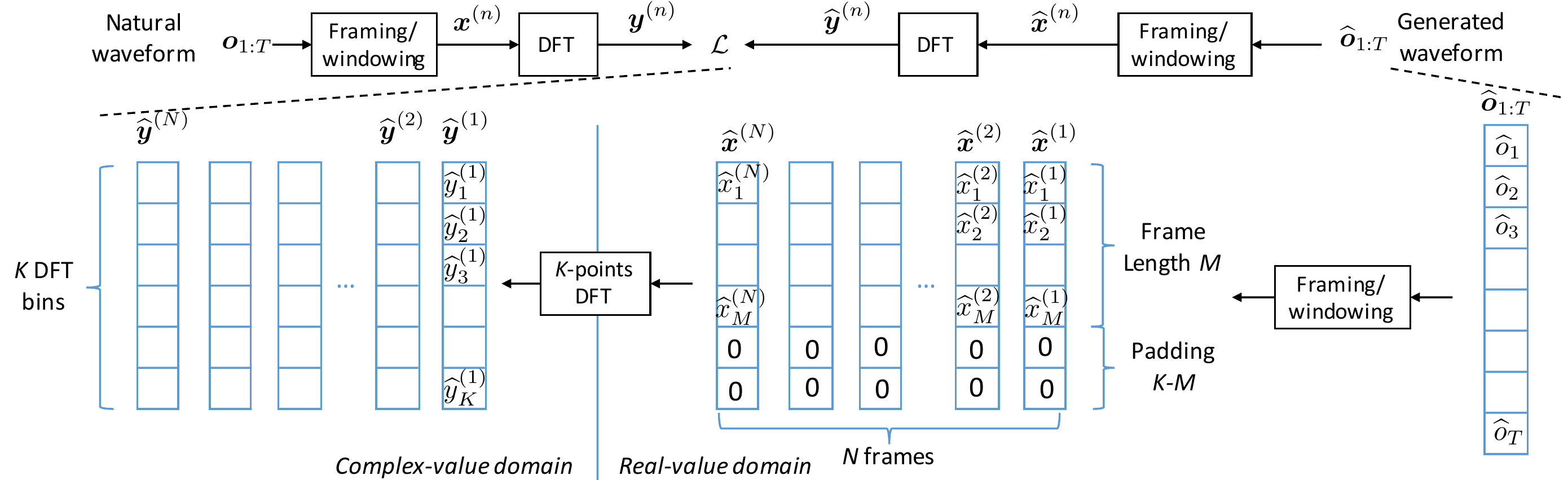}}
\caption{Framing/windowing and DFT steps. $T$, $M$, $K$ denotes waveform length, frame length, and number of DFT bins.}
\label{fig_append_1}
\end{figure*}

Figure~\ref{fig_append_1} plots the two steps to derive the spectrum from the generated waveform $\widehat{\bs{o}}_{1:T}$.
We use $\widehat{\bs{x}}^{(n)}=[\widehat{x}^{(n)}_1, \cdots, \widehat{x}^{(n)}_M]^{\top}\in\mathbb{R}^{M}$ to denote the $n$-th waveform frame of length $M$. We then use $\widehat{\bs{y}}^{(n)}=[\widehat{y}^{(n)}_1, \cdots, \widehat{y}^{(n)}_K]^{\top}\in\mathbb{C}^{K}$ to denote the spectrum of $\widehat{\bs{x}}^{(n)}$ calculated using $K$-point DFT, i.e., $\widehat{\bs{y}}^{(n)}=\text{DFT}_{K}(\widehat{\bs{x}}^{(n)})$. 
For fast Fourier transform, $K$ is set to the power of 2, and $\widehat{\bs{x}}^{(n)}$ is zero-padded to length $K$ before DFT.

\subsection{Framing and windowing}
\begin{figure*}[h]
\centering
\fbox{\includegraphics[width=0.95\textwidth]{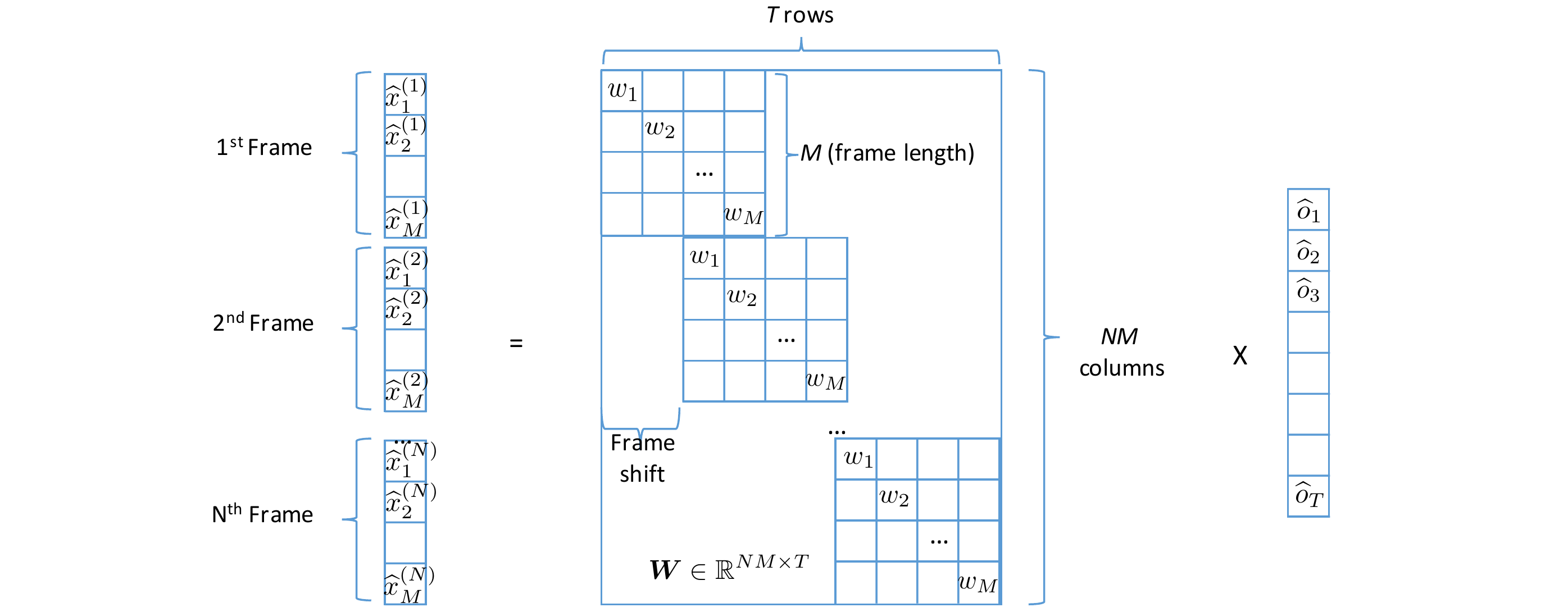}}
\caption{A matrix format of framing/windowing operation, where $\bs{w}_{1:M}$ denote coefficients of Hann window.}
\label{fig_append_2}
\end{figure*}

The framing and windowing operation is also parallelized over $\widehat{x}^{(n)}_{m}$ using \texttt{for\_each} command in CUDA/Thrust. 
However, for explanation, let's use the matrix operation in Figure~\ref{fig_append_2}. In other words, we compute
\begin{equation}
\widehat{x}^{(n)}_{m} = \sum_{t=1}^{T}\widehat{o}_tw^{(n,m)}_t, 
\label{eq:append_fw}
\end{equation}
where $w^{(n,m)}_t$ is the element in the $\big[(n-1)\times{M}+m\big]$-th row and the $t$-th column of the transformation matrix $\bs{W}$. 

\subsection{DFT}
Our implementation uses cuFFT (\texttt{cufftExecR2C} and \texttt{cufftPlan1d}) \footnote{https://docs.nvidia.com/cuda/cufft/index.html} to compute $\{\bs{y}^{(1)}, \cdots, \bs{y}^{(N)}\}$ in parallel. 

\section{Backward computation}
\begin{figure*}[h]
\centering
\fbox{\includegraphics[width=\textwidth]{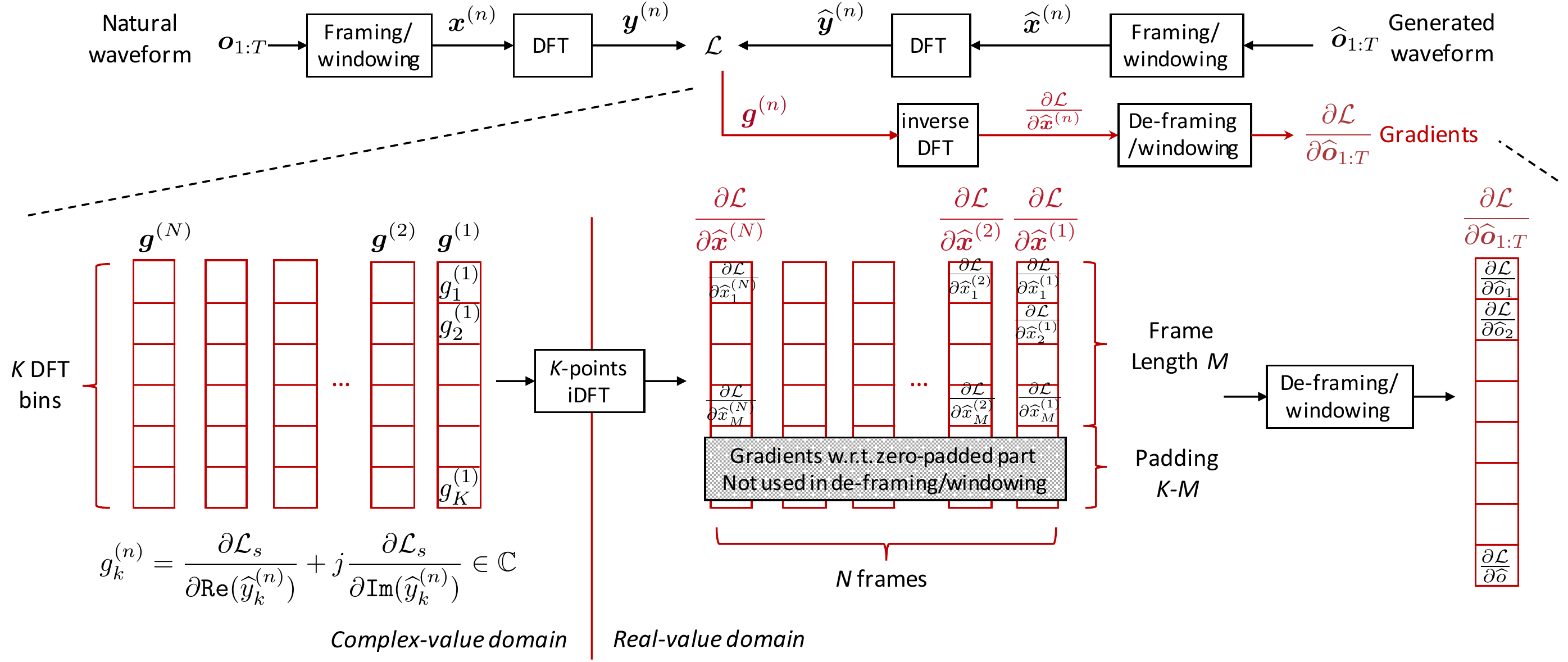}}
\vspace{-6mm}
\caption{Steps to compute gradients}
\label{fig_append_3}
\end{figure*}

For back-propagation, we need to compute the gradient $\frac{\partial{\mathcal{L}}}{\partial{\widehat{\bs{o}}_{1:T}}}\in\mathbb{R}^{T}$ following the steps plotted in Figure~\ref{fig_append_3}.

\subsection{The 2nd step: from $\frac{\partial\mathcal{L}}{\partial\widehat{{x}}^{(n)}_m}$ to $\frac{\partial{\mathcal{L}}}{\partial{\widehat{{o}}_{t}}}$}

Suppose we have $\{\frac{\partial\mathcal{L}}{\partial\widehat{\bs{x}}^{(1)}}, \cdots, \frac{\partial\mathcal{L}}{\partial\widehat{\bs{x}}^{(N)}}\}$, where each $\frac{\partial\mathcal{L}}{\partial\widehat{\bs{x}}^{(n)}}\in\mathbb{R}^{M}$ and $\frac{\partial\mathcal{L}}{\partial\widehat{{x}}^{(n)}_{m}}\in\mathbb{R}$. Then,
we can compute $\frac{\partial{\mathcal{L}}}{\partial{\widehat{{o}}_{t}}}$ on the basis Equation~(\ref{eq:append_fw}) as
\begin{equation}
\frac{\partial{\mathcal{L}}}{\partial{\widehat{{o}}_{t}}} = \sum_{n=1}^{N}\sum_{m=1}^{M}\frac{\partial\mathcal{L}}{\partial\widehat{{x}}^{(n)}_{m}}w_{t}^{(n,m)},
\label{eq:append_deframe}
\end{equation}
where $w_{t}^{(n,m)}$ are the framing/windowing coefficients. This equation explains what we mean by saying `$\frac{\partial{\mathcal{L}}}{\partial{\widehat{{o}}_{t}}}$ can be easily accumulated the relationship between $\widehat{{o}}_t$ and each $\widehat{x}^{(n)}_{m}$ has been determined by the framing and windowing operations'. 

Our implementation uses CUDA/Thrust \texttt{for\_each} command to launch $T$ threads and compute $\frac{\partial{\mathcal{L}}}{\partial{\widehat{{o}}_{t}}}, t\in\{1,\cdots,T\}$ in parallel. 
Because $\widehat{o}_t$ is only used in a few frames and there is only one $w_{t}^{(n,m)}\neq0$ for each $\{n, t\}$, Equation~(\ref{eq:append_deframe}) can be optimized as
\begin{equation}
\frac{\partial{\mathcal{L}}}{\partial{\widehat{{o}}_{t}}} = \sum_{n=N_{t,min}}^{N_{t,max}}\frac{\partial\mathcal{L}}{\partial\widehat{{x}}^{(n)}_{m_{t,n}}}w_{t}^{(n,m_{t,n})},
\label{eq:append_deframe_2}
\end{equation}
where $[N_{t,min}, N_{t,max}]$ is the frame range that $\widehat{o}_t$ appears, and $m_{t,n}$ is the position of $\widehat{o}_t$ in the $n$-th frame.

\subsection{The 1st step: compute $\frac{\partial\mathcal{L}}{\partial\widehat{{x}}^{(n)}_m}$}
Remember that $\widehat{\bs{y}}^{(n)}=[\widehat{y}^{(n)}_1, \cdots, \widehat{y}^{(n)}_K]^{\top}\in\mathbb{C}^{K}$ is the K-points DFT spectrum of $\widehat{\bs{x}}^{(n)}=[\widehat{x}^{(n)}_1, \cdots, \widehat{x}^{(n)}_M]^{\top}\in\mathbb{R}^{M}$. Therefore we know
\begin{align}
\texttt{Re}(\widehat{{y}}^{(n)}_k) &= \sum_{m=1}^{M} \widehat{x}^{(n)}_m \cos(\frac{2\pi}{K}(k-1)(m-1)), \\
\texttt{Im}(\widehat{{y}}^{(n)}_k) &= -\sum_{m=1}^{M} \widehat{x}^{(n)}_m \sin(\frac{2\pi}{K}(k-1)(m-1)),
\end{align}
where $k\in[1, K]$. Note that, although the sum should be $\sum_{m=1}^{K}$,  the summation over the zero-padded part $\sum_{m=M+1}^{K}0\cos(\frac{2\pi}{K}(k-1)(m-1))$ can be safely ignored \footnote{Although we can avoid zero-padding by setting $K=M$, in practice $K$ is usually the power of 2 while the frame length $M$ is not.}. 

Suppose we compute a log spectral amplitude distance $\mathcal{L}$ over the $N$ frames as
\begin{equation}
\mathcal{L} = \frac{1}{2}\sum_{n=1}^{N}\sum_{k=1}^{K}\Big[\log\frac{\texttt{Re}(y_k^{(n)})^2+\texttt{Im}(y_k^{(n)})^2}{\texttt{Re}(\widehat{y}_k^{(n)})^2+\texttt{Im}(\widehat{y}_k^{(n)})^2}\Big]^2.
\label{eq:apped_dft_spectral}
\end{equation}
Because $\mathcal{L}$, $\widehat{{x}}^{(n)}_m$, $\texttt{Re}(\widehat{{y}}^{(n)}_k)$, and $\texttt{Im}(\widehat{{y}}^{(n)}_k)$ are real-valued numbers, we can compute the gradient $\frac{\partial\mathcal{L}}{\partial\widehat{{x}}^{(n)}_m}$ using the chain rule:
\begin{align}
\frac{\partial\mathcal{L}}{\partial\widehat{{x}}^{(n)}_m} &= 
\sum_{k=1}^{K}\frac{\partial\mathcal{L}}{\partial\texttt{Re}(\widehat{y}_k^{(n)})}\frac{\partial\texttt{Re}(\widehat{y}_k^{(n)})}{\partial\widehat{{x}}^{(n)}_m} + \sum_{k=1}^{K}\frac{\partial\mathcal{L}}{\partial\texttt{Im}(\widehat{y}_k^{(n)})}\frac{\partial\texttt{Im}(\widehat{y}_k^{(n)})}{\partial\widehat{{x}}^{(n)}_m} 
\label{eq:apped_dft_spectral_grad_1}\\
&= \sum_{k=1}^{K}\frac{\partial\mathcal{L}}{\partial\texttt{Re}(\widehat{y}_k^{(n)})}\cos(\frac{2\pi}{K}(k-1)(m-1)) -\sum_{k=1}^{K}\frac{\partial\mathcal{L}}{\partial\texttt{Im}(\widehat{y}_k^{(n)})}\sin(\frac{2\pi}{K}(k-1)(m-1)).
\label{eq:apped_dft_spectral_grad_2}
\end{align}
Once we compute $\frac{\partial\mathcal{L}}{\partial\widehat{{x}}^{(n)}_m}$ for each $m$ and $n$, we can use Equation~(\ref{eq:append_deframe_2}) to compute the gradient $\frac{\partial{\mathcal{L}}}{\partial{\widehat{{o}}_{t}}}$.

\subsection{Implementation of the 1st step using inverse DFT}
Because $\frac{\partial\mathcal{L}}{\partial\texttt{Re}(\widehat{y}_k^{(n)})}$ and $\frac{\partial\mathcal{L}}{\partial\texttt{Im}(\widehat{y}_k^{(n)})}$ are real numbers, we can directly implement Equation~(\ref{eq:apped_dft_spectral_grad_2}) using matrix multiplication. 
However, a more efficient way is to use inverse DFT (iDFT).

Suppose a complex valued signal 
$\bs{g}=[{g}_1, {g}_2, \cdots, {g}_L]\in\mathbb{C}^{K}$, we compute $\bs{b}=[b_1, \cdots, b_K]$ as the K-point inverse DFT of $\bs{g}$ by \footnote{
cuFFT performs un-normalized FFTs, i.e., the scaling factor $\frac{1}{K}$ is not used.}
\begin{align}
b_m = &\sum_{k=1}^{K} g_k e^{j\frac{2\pi}{K}(k-1)(m-1)} 
\label{eq:apped_idft_1}\\
= &\sum_{k=1}^{K} [\texttt{Re}(g_k) + j\texttt{Im}(g_k)] [\cos(\frac{2\pi}{K}(k-1)(m-1)) + j\sin(\frac{2\pi}{K}(k-1)(m-1))] 
\label{eq:apped_idft_2}\\
= &\sum_{k=1}^{K} \texttt{Re}(g_k) \cos(\frac{2\pi}{K}(k-1)(m-1)) - \sum_{k=1}^{K} \texttt{Im}(g_k)\sin(\frac{2\pi}{K}(k-1)(m-1)) 
\label{eq:apped_idft_3}\\
&+ j\Big[\sum_{k=1}^{K} \texttt{Re}(g_k)\sin(\frac{2\pi}{K}(k-1)(m-1)) +  \sum_{k=1}^{K} \texttt{Im}(g_k)\cos(\frac{2\pi}{K}(k-1)(m-1)) \Big].
\label{eq:apped_idft_4}
\end{align}
For the first term in Line~\ref{eq:apped_idft_4}, we can write
\begin{align}
&\sum_{k=1}^{K} \texttt{Re}(g_k)\sin(\frac{2\pi}{K}(k-1)(m-1)) \\
=&\texttt{Re}(g_0)\sin(\frac{2\pi}{K}(1-1)(m-1)) + \texttt{Re}(g_{\frac{K}{2}+1})\sin(\frac{2\pi}{K}(\frac{K}{2}+1-1)(m-1)) 
\label{eq:apped_idft_5} \\
&+ \sum_{k=2}^{\frac{K}{2}} \texttt{Re}(g_k)\sin(\frac{2\pi}{K}(k-1)(m-1)) + \sum_{k=\frac{K}{2}+2}^{K} \texttt{Re}(g_k)\sin(\frac{2\pi}{K}(k-1)(m-1))   
\label{eq:apped_idft_6}\\
=&\sum_{k=2}^{\frac{K}{2}} \Big[\texttt{Re}(g_k)\sin(\frac{2\pi}{K}(k-1)(m-1)) + \texttt{Re}(g_{(K+2-k}))\sin(\frac{2\pi}{K}({K}+2-k-1)(m-1))\Big] 
\label{eq:apped_idft_7}\\
=&\sum_{k=2}^{\frac{K}{2}} \Big[\texttt{Re}(g_k) - \texttt{Re}(g_{(K+2-k}))\Big]\sin(\frac{2\pi}{K}(k-1)(m-1))
\label{eq:apped_idft_8}
\end{align}
Note that in Line~(\ref{eq:apped_idft_5}), $\texttt{Re}(g_1)\sin(\frac{2\pi}{K}(1-1)(m-1)) = \texttt{Re}(g_1)\sin(0)= 0$, and $\texttt{Re}(g_{\frac{K}{2}+1})\sin(\frac{2\pi}{K}(\frac{K}{2}+1-1)(m-1)) = \texttt{Re}(g_{\frac{K}{2}+1})\sin((m-1)\pi) = 0$. 

It is easy to those that Line~(\ref{eq:apped_idft_8}) is equal to 0 if $\texttt{Re}(g_k)=\texttt{Re}(g_{(K+2-k)})$, for any $k\in[2, \frac{K}{2}]$. 
Similarly, it can be shown that $\sum_{k=1}^{K} \texttt{Im}(g_k)\cos(\frac{2\pi}{K}(k-1)(m-1))=0$ if 
$\texttt{Im}(g_k)=-\texttt{Im}(g_{(K+2-k)}), k\in[2, \frac{K}{2}]$ and $\texttt{Im}(g_1)=\texttt{Im}(g_{(\frac{K}{2}+1)})=0$.
When these two terms are equal to 0, the imaginary part in Line~(\ref{eq:apped_idft_4}) will be 0, and $b_m = \sum_{k=1}^{K} g_k e^{j\frac{2\pi}{K}(k-1)(m-1)}$ in Line~(\ref{eq:apped_idft_1}) will be a real number.

To summarize, if $\bs{g}$ satisfies the conditions below
\begin{align}
\texttt{Re}(g_k) &=\texttt{Re}(g_{(K+2-k)}), \quad{}k\in[2, \frac{K}{2}] \label{eq:apped_conjugate_sym_1}\\
\texttt{Im}(g_k) &=
\begin{cases}
-\texttt{Im}(g_{(K+2-k)}), \qquad{}k\in[2, \frac{K}{2}] \\
0, \qquad\qquad\qquad\qquad{}k=\{1, \frac{K}{2}+1\}
\end{cases}
\label{eq:apped_conjugate_sym_2}
\end{align}
inverse DFT of $\bs{g}$ will be real-valued:
\begin{equation}
\sum_{k=1}^{K} g_k e^{j\frac{2\pi}{K}(k-1)(m-1)}  = \sum_{k=1}^{K} \texttt{Re}(g_k) \cos(\frac{2\pi}{K}(k-1)(m-1)) - \sum_{k=1}^{K} \texttt{Im}(g_k)\sin(\frac{2\pi}{K}(k-1)(m-1)) 
\label{eq:apped_idft_exact}
\end{equation}
This is a basic concept in signal processing: the iDFT of a conjugate-symmetric (Hermitian)\footnote{It should be called circular conjugate symmetry in strict sense} signal will be a real-valued signal. 

We can observer from Equation~(\ref{eq:apped_idft_exact}) and (\ref{eq:apped_dft_spectral_grad_2}) that, if $\Big[\frac{\partial\mathcal{L}}{\partial\texttt{Re}(\widehat{y}_1^{(n)})} + j\frac{\partial\mathcal{L}}{\partial\texttt{Im}(\widehat{y}_1^{(n)})}, \cdots, \frac{\partial\mathcal{L}}{\partial\texttt{Re}(\widehat{y}_K^{(n)})} + j\frac{\partial\mathcal{L}}{\partial\texttt{Im}(\widehat{y}_K^{(n)})}\Big]^{\top}$ is conjugate-symmetric, the gradient vector $\frac{\partial\mathcal{L}}{\partial\widehat{\bs{x}}^{(n)}} = [\frac{\partial\mathcal{L}}{\partial\widehat{{x}}^{(n)}_1},\cdots,\frac{\partial\mathcal{L}}{\partial\widehat{{x}}^{(n)}_M}]^{\top}$ be computed using iDFT:
\begin{equation}
\begin{bmatrix}
\frac{\partial\mathcal{L}}{\partial\widehat{{x}}^{(n)}_1} \\[1em]
\frac{\partial\mathcal{L}}{\partial\widehat{{x}}^{(n)}_2} \\[1em]
\cdots \\[1em]
\frac{\partial\mathcal{L}}{\partial\widehat{{x}}^{(n)}_M} \\[1em]
\frac{\partial\mathcal{L}}{\partial\widehat{{x}}^{(n)}_{M+1}} \\[1em]
\cdots \\[1em]
\frac{\partial\mathcal{L}}{\partial\widehat{{x}}^{(n)}_{K}}
\end{bmatrix} 
= \text{iDFT}(
\begin{bmatrix}
\frac{\partial\mathcal{L}}{\partial\texttt{Re}(\widehat{y}_1^{(n)})} + j\frac{\partial\mathcal{L}}{\partial\texttt{Im}(\widehat{y}_1^{(n)})} \\[1em]
\frac{\partial\mathcal{L}}{\partial\texttt{Re}(\widehat{y}_2^{(n)})} + j\frac{\partial\mathcal{L}}{\partial\texttt{Im}(\widehat{y}_2^{(n)})} \\[1em]
\cdots \\[1em]
\frac{\partial\mathcal{L}}{\partial\texttt{Re}(\widehat{y}_M^{(n)})} + j\frac{\partial\mathcal{L}}{\partial\texttt{Im}(\widehat{y}_M^{(n)})} \\[1em]
\frac{\partial\mathcal{L}}{\partial\texttt{Re}(\widehat{y}_{M+1}^{(n)})} + j\frac{\partial\mathcal{L}}{\partial\texttt{Im}(\widehat{y}_{M+1}^{(n)})} \\[1em]
\cdots \\[1em]
\frac{\partial\mathcal{L}}{\partial\texttt{Re}(\widehat{y}_K^{(n)})} + j\frac{\partial\mathcal{L}}{\partial\texttt{Im}(\widehat{y}_K^{(n)})} 
\end{bmatrix} 
).
\end{equation}
Note that $\{\frac{\partial\mathcal{L}}{\partial\widehat{{x}}^{(n)}_{M+1}}, \cdots, \frac{\partial\mathcal{L}}{\partial\widehat{{x}}^{(n)}_{K}}\}$ are the gradients w.r.t to the zero-padded part, which will not be used and can safely set to 0. 
The iDFT of a conjugate symmetric signal can be executed using cuFFT \texttt{cufftExecC2R} command. It is more efficient than other implementations of Equation~(\ref{eq:apped_dft_spectral_grad_2}) because
\begin{itemize}
\item there is no need to compute the imaginary part;
\item there is no need to compute and allocate GPU memory for $g_k$ where $k\in[\frac{K}{2}+2, K]$ because of the conjugate symmetry;
\item iDFT can be executed for all the $N$ frames in parallel.
\end{itemize}

\subsection{Conjugate symmetry complex-valued gradient vector}
The conjugate symmetry of $\Big[\frac{\partial\mathcal{L}}{\partial\texttt{Re}(\widehat{y}_1^{(n)})} + j\frac{\partial\mathcal{L}}{\partial\texttt{Im}(\widehat{y}_1^{(n)})}, \cdots, \frac{\partial\mathcal{L}}{\partial\texttt{Re}(\widehat{y}_K^{(n)})} + j\frac{\partial\mathcal{L}}{\partial\texttt{Im}(\widehat{y}_K^{(n)})}\Big]^{\top}$ is satisfied if $\mathcal{L}$ is carefully chosen. Luckily, most of the common distance metrics can be used.

\subsubsection{Log spectral amplitude distance}
Given the log spectral amplitude distance $\mathcal{L}_s$ in Equation~(\ref{eq:apped_dft_spectral}), we can compute
\begin{align}
\frac{\partial\mathcal{L}_s}{\partial\texttt{Re}(\widehat{y}_k^{(n)})} &= 
\Big[\log[\texttt{Re}(\widehat{y}_k^{(n)})^2 + {\texttt{Im}(\widehat{y}_k^{(n)}})^2] - \log[\texttt{Re}(y_k^{(n)})^2 + \texttt{Im}(y_k^{(n)})^2]\Big]\frac{2\texttt{Re}(\widehat{y}_k^{(n)})}{\texttt{Re}(\widehat{y}_k^{(n)})^2 + {\texttt{Im}(\widehat{y}_k^{(n)}})^2} \label{eq:apped_idft_log_amp_g_1} \\
\frac{\partial\mathcal{L}_s}{\partial\texttt{Im}(\widehat{y}_k^{(n)})} &= 
\Big[\log[\texttt{Re}(\widehat{y}_k^{(n)})^2 + {\texttt{Im}(\widehat{y}_k^{(n)}})^2] - \log[\texttt{Re}(y_k^{(n)})^2 + \texttt{Im}(y_k^{(n)})^2]\Big]\frac{2\texttt{Im}(\widehat{y}_k^{(n)})}{\texttt{Re}(\widehat{y}_k^{(n)})^2 + {\texttt{Im}(\widehat{y}_k^{(n)}})^2} \label{eq:apped_idft_log_amp_g_2}
\end{align}
Because $\widehat{\bs{y}}^{(n)}$ is the DFT spectrum of the real-valued signal, $\widehat{\bs{y}}^{(n)}$ is conjugate symmetric, and $\texttt{Re}(\widehat{y}_k^{(n)})$ and  $\texttt{Im}(\widehat{y}_k^{(n)})$ satisfy the condition in Equations~(\ref{eq:apped_conjugate_sym_1}) and (\ref{eq:apped_conjugate_sym_2}), respectively.
Because the amplitude $\texttt{Re}(\widehat{y}_k^{(n)})^2 + {\texttt{Im}(\widehat{y}_k^{(n)}})^2$ does not change the symmetry, $\frac{\partial\mathcal{L}_s}{\partial\texttt{Re}(\widehat{y}_k^{(n)})}$ and $\frac{\partial\mathcal{L}_s}{\partial\texttt{Im}(\widehat{y}_k^{(n)})}$ also satisfy the conditions in Equations~(\ref{eq:apped_conjugate_sym_1}) and (\ref{eq:apped_conjugate_sym_2}), respectively, and $\Big[\frac{\partial\mathcal{L}_s}{\partial\texttt{Re}(\widehat{y}_1^{(n)})} + j\frac{\partial\mathcal{L}_s}{\partial\texttt{Im}(\widehat{y}_1^{(n)})}, \cdots, \frac{\partial\mathcal{L}_s}{\partial\texttt{Re}(\widehat{y}_K^{(n)})} + j\frac{\partial\mathcal{L}_s}{\partial\texttt{Im}(\widehat{y}_K^{(n)})}\Big]^{\top}$ is conjugate-symmetric.

\subsubsection{Phase distance}
Let $\widehat\theta^{(n)}_k$ and $\theta^{(n)}_k$ to be the phases of $\widehat{y}^{(n)}_k$ and ${y}^{(n)}_k$, respectively. Then, the phase distance is defined as 
\begin{align}
\mathcal{L}_p &= \frac{1}{2}\sum_{n=1}^{N}\sum_{k=1}^{K}\Big|1-\exp(j(\widehat\theta^{(n)}_k - {\theta}^{(n)}_k))\Big|^2\\
&= \frac{1}{2}\sum_{n=1}^{N}\sum_{k=1}^{K}\Big|1-\cos(\widehat\theta^{(n)}_k - {\theta}^{(n)}_k)) - j\sin(\widehat\theta^{(n)}_k - {\theta}^{(n)}_k))\Big|^2\\
&= \frac{1}{2}\sum_{n=1}^{N}\sum_{k=1}^{K}\Big[\big(1-\cos(\widehat\theta^{(n)}_k - {\theta}^{(n)}_k))\big)^2 + \sin(\widehat\theta^{(n)}_k - {\theta}^{(n)}_k))^2\Big]\\
&= \frac{1}{2}\sum_{n=1}^{N}\sum_{k=1}^{K}\Big[1 + \cos(\widehat\theta^{(n)}_k - {\theta}^{(n)}_k))^2 + \sin(\widehat\theta^{(n)}_k - {\theta}^{(n)}_k))^2 -2\cos(\widehat\theta^{(n)}_k - {\theta}^{(n)}_k)) \Big]\\
&= \sum_{n=1}^{N}\sum_{k=1}^{K}\Big(1 -\cos(\widehat\theta^{(n)}_k - {\theta}^{(n)}_k)) \Big)\\
&= \sum_{n=1}^{N}\sum_{k=1}^{K}\Big[1 -\big(\cos(\widehat\theta^{(n)}_k)\cos({\theta}^{(n)}_k)) + \sin(\widehat\theta^{(n)}_k)\sin({\theta}^{(n)}_k))\big)\Big]\\
&=\sum_{n=1}^{N}\sum_{k=1}^{K}\Big[1-\frac{\texttt{Re}(\widehat{y}_k^{(n)})\texttt{Re}({y}_k^{(n)})+\texttt{Im}(\widehat{y}_k^{(n)})\texttt{Im}({y}_k^{(n)})}{\sqrt{\texttt{Re}(\widehat{y}_k^{(n)})^2 + \texttt{Im}(\widehat{y}_k^{(n)})^2}\sqrt{\texttt{Re}({y}_k^{(n)})^2 + \texttt{Im}({y}_k^{(n)})^2}}\Big],
\label{eq:append_dft_phase}
\end{align}
where 
\begin{align}
\cos(\widehat{\theta}_{k}^{(n)}) &= \frac{\texttt{Re}(\widehat{y}_k^{(n)})}{\sqrt{\texttt{Re}(\widehat{y}_k^{(n)})^2 + \texttt{Im}(\widehat{y}_k^{(n)})^2}},\quad \cos({\theta}_{k}^{(n)}) = \frac{\texttt{Re}({y}_k^{(n)})}{\sqrt{\texttt{Re}({y}_k^{(n)})^2 + \texttt{Im}({y}_k^{(n)})^2}} \\
\sin(\widehat{\theta}_{k}^{(n)}) &= \frac{\texttt{Im}(\widehat{y}_k^{(n)})}{\sqrt{\texttt{Re}(\widehat{y}_k^{(n)})^2 + \texttt{Im}(\widehat{y}_k^{(n)})^2}},\quad \sin({\theta}_{k}^{(n)}) = \frac{\texttt{Im}({y}_k^{(n)})}{\sqrt{\texttt{Re}({y}_k^{(n)})^2 + \texttt{Im}({y}_k^{(n)})^2}}.
\end{align}
Therefore, we get 
\begin{align}
\frac{\partial{\mathcal{L}_p}}{\partial\texttt{Re}(\widehat{y}_k^{(n)})} =& -\cos(\theta_{k}^{(n)})\frac{\partial{\cos(\widehat{\theta}_k^{(n)})}}{\partial\texttt{Re}(\widehat{y}_k^{(n)})} - \sin(\theta_{k}^{(n)})\frac{\partial{\sin(\widehat{\theta}_k^{(n)})}}{\partial\texttt{Re}(\widehat{y}_k^{(n)})} \\ 
=& -\cos(\theta_{k}^{(n)})
\frac{\sqrt{\texttt{Re}(\widehat{y}_k^{(n)})^2 + \texttt{Im}(\widehat{y}_k^{(n)})^2} - \texttt{Re}(\widehat{y}_k^{(n)})\frac{1}{2}\frac{2\texttt{Re}(\widehat{y}_k^{(n)})}{\sqrt{\texttt{Re}(\widehat{y}_k^{(n)})^2 + \texttt{Im}(\widehat{y}_k^{(n)})^2}}}{{\texttt{Re}(\widehat{y}_k^{(n)})^2 + \texttt{Im}(\widehat{y}_k^{(n)})^2}}- \sin(\theta_{k}^{(n)})\frac{-\texttt{Im}(\widehat{y}_k^{(n)})\frac{1}{2}\frac{2\texttt{Re}(\widehat{y}_k^{(n)})}{\sqrt{\texttt{Re}(\widehat{y}_k^{(n)})^2 + \texttt{Im}(\widehat{y}_k^{(n)})^2}}}{{\texttt{Re}(\widehat{y}_k^{(n)})^2 + \texttt{Im}(\widehat{y}_k^{(n)})^2}} \\ 
=& -\cos(\theta_{k}^{(n)}) 
\frac{\texttt{Re}(\widehat{y}_k^{(n)})^2 + \texttt{Im}(\widehat{y}_k^{(n)})^2 - \texttt{Re}(\widehat{y}_k^{(n)})^2}{\big({\texttt{Re}(\widehat{y}_k^{(n)})^2 + \texttt{Im}(\widehat{y}_k^{(n)})^2}\big)^{\frac{3}{2}}}
- \sin(\theta_{k}^{(n)})\frac{-\texttt{Im}(\widehat{y}_k^{(n)})\texttt{Re}(\widehat{y}_k^{(n)})}{\big({\texttt{Re}(\widehat{y}_k^{(n)})^2 + \texttt{Im}(\widehat{y}_k^{(n)})^2}\big)^{\frac{3}{2}}}\\
=& -
\frac{\texttt{Re}({y}_k^{(n)})\texttt{Re}(\widehat{y}_k^{(n)})^2 + \texttt{Re}({y}_k^{(n)})\texttt{Im}(\widehat{y}_k^{(n)})^2 -\texttt{Re}({y}_k^{(n)})\texttt{Re}(\widehat{y}_k^{(n)})^2 - \texttt{Im}({y}_k^{(n)})\texttt{Im}(\widehat{y}_k^{(n)})\texttt{Re}(\widehat{y}_k^{(n)})}{
\big({\texttt{Re}({y}_k^{(n)})^2 + \texttt{Im}({y}_k^{(n)})^2}\big)^{\frac{1}{2}}\big({\texttt{Re}(\widehat{y}_k^{(n)})^2 + \texttt{Im}(\widehat{y}_k^{(n)})^2}\big)^{\frac{3}{2}}} \\
=& -
\frac{\texttt{Re}({y}_k^{(n)})\texttt{Im}(\widehat{y}_k^{(n)}) - \texttt{Im}({y}_k^{(n)})\texttt{Re}(\widehat{y}_k^{(n)})}{
\big({\texttt{Re}({y}_k^{(n)})^2 + \texttt{Im}({y}_k^{(n)})^2}\big)^{\frac{1}{2}}\big({\texttt{Re}(\widehat{y}_k^{(n)})^2 + \texttt{Im}(\widehat{y}_k^{(n)})^2}\big)^{\frac{3}{2}}}\texttt{Im}(\widehat{y}_k^{(n)}) \\
\frac{\partial{\mathcal{L}_p}}{\partial\texttt{Im}(\widehat{y}_k^{(n)})} =& -\cos(\theta_{k}^{(n)})\frac{\partial{\cos(\widehat{\theta}_k^{(n)})}}{\partial\texttt{Im}(\widehat{y}_k^{(n)})} - \sin(\theta_{k}^{(n)})\frac{\partial{\sin(\widehat{\theta}_k^{(n)})}}{\partial\texttt{Im}(\widehat{y}_k^{(n)})} \\ 
=& -
\frac{\texttt{Im}({y}_k^{(n)})\texttt{Re}(\widehat{y}_k^{(n)}) - \texttt{Re}({y}_k^{(n)})\texttt{Im}(\widehat{y}_k^{(n)})}{
\big({\texttt{Re}({y}_k^{(n)})^2 + \texttt{Im}({y}_k^{(n)})^2}\big)^{\frac{1}{2}}\big({\texttt{Re}(\widehat{y}_k^{(n)})^2 + \texttt{Im}(\widehat{y}_k^{(n)})^2}\big)^{\frac{3}{2}}}\texttt{Re}(\widehat{y}_k^{(n)}).
\end{align}

Because both $\bs{y}^{(n)}$ and $\bs{\widehat{y}}^{(n)}$ are conjugate-symmetric, it can be easily observed that $\frac{\partial{\mathcal{L}_p}}{\partial\texttt{Re}(\widehat{y}_k^{(n)})}$ and $\frac{\partial{\mathcal{L}_p}}{\partial\texttt{Re}(\widehat{y}_k^{(n)})}$ satisfy the condition in  Equations~(\ref{eq:apped_conjugate_sym_1}) and (\ref{eq:apped_conjugate_sym_2}), respectively.

\section{Multiple distance metrics}
Different distance metrics can be merged easily. For example, we can define 
\begin{equation}
\mathcal{L} = \mathcal{L}_{s1} + \cdots + \mathcal{L}_{sS} + \mathcal{L}_{p1} + \cdots + \mathcal{L}_{pP},
\end{equation}
where $\mathcal{L}_{s*}\in\mathbb{R}$  and $\mathcal{L}_{p*}\in\mathbb{R}$ may use different numbers of DFT bins, frame length, or frame shift. 
Although the dimension of the gradient vector $\frac{\partial\mathcal{L}_{*}}{\partial\widehat{\bs{x}}^{(n)}}$ may be different, the gradient $\frac{\partial\mathcal{L}_{*}}{\partial{\widehat{\bs{o}}_{1:T}}}\in\mathbb{R}^{T}$ will always be 
a real-valued vector of dimension $T$ after de-framing/windowing. The gradients then will be simply merged together as
\begin{equation}
\frac{\partial\mathcal{L}}{\partial{\widehat{\bs{o}}_{1:T}}} = \frac{\partial\mathcal{L}_{s1}}{\partial{\widehat{\bs{o}}_{1:T}}} + 
\cdots + \frac{\partial\mathcal{L}_{sS}}{\partial{\widehat{\bs{o}}_{1:T}}} + 
\frac{\partial\mathcal{L}_{p1}}{\partial{\widehat{\bs{o}}_{1:T}}} + \cdots + 
\frac{\partial\mathcal{L}_{pP}}{\partial{\widehat{\bs{o}}_{1:T}}}.
\end{equation}

\begin{figure*}[h]
\centering
\fbox{\includegraphics[width=\textwidth]{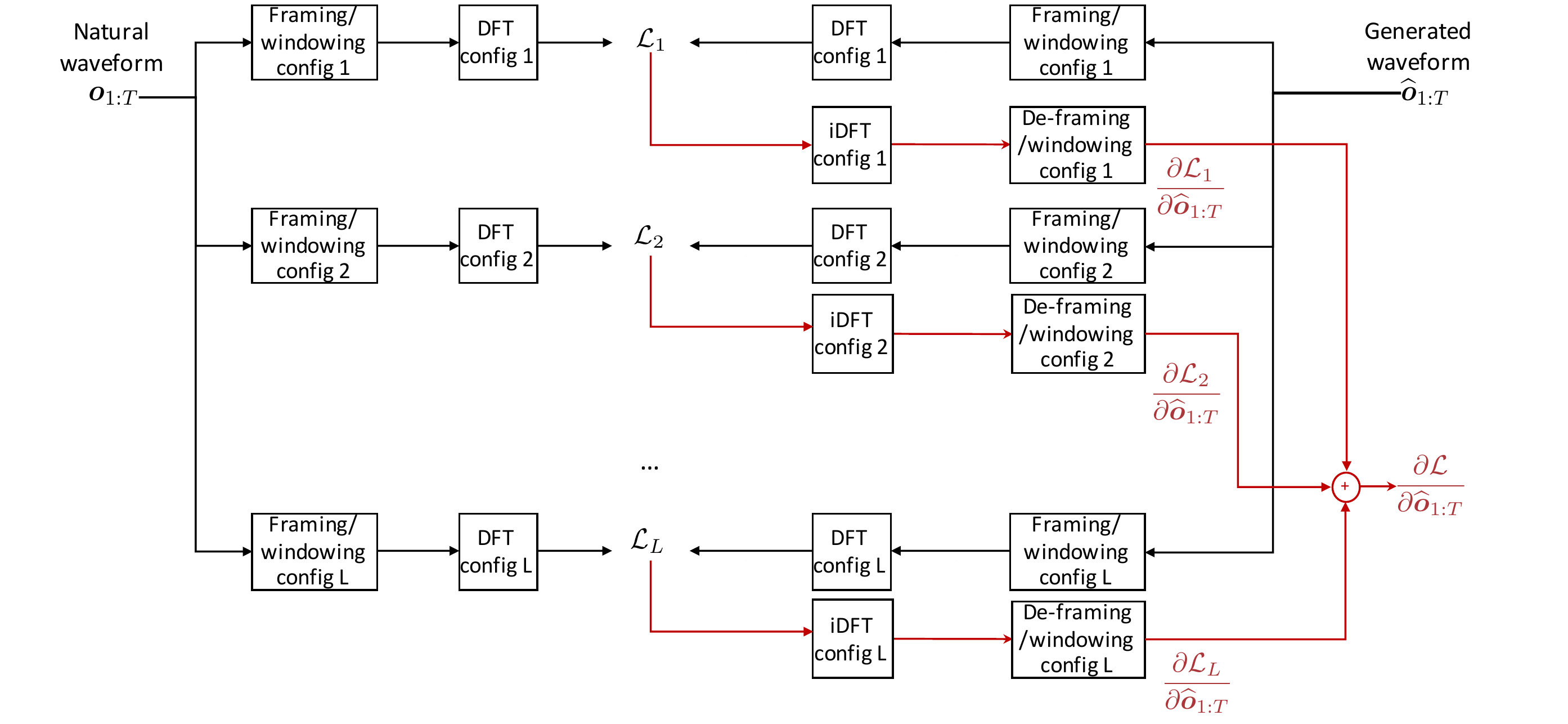}}
\vspace{-6mm}
\caption{Using multiple distances $\{\mathcal{L}_1, \cdots, \mathcal{L}_L\}$}
\label{fig_append_4}
\end{figure*}

\end{document}